\documentclass[12pt,epsfig,multirow]{article}
\usepackage{graphicx}
\usepackage[usenames]{color}
\newcommand{\be}{\begin{eqnarray}}
\newcommand{\ee}{\end{eqnarray}}

\def\lan{\left\langle}
\def\ran{\right\rangle}
\def\dis{\displaystyle}
\def\barr{\begin{array}}
\def\earr{\end{array}}

\begin{document}

\thispagestyle{empty}
\begin{flushright}

\end{flushright}
\bigskip

\begin{center}
{\Large \bf Statistical Spectroscopy for Neutron-rich sd-Shell Nuclei} 

\vspace{.5in}

{\bf {Kamales Kar$^{\star, a}$}} 
\vskip .5cm
$^\star${\normalsize \it Saha Institute of Nuclear Physics,}\\
{\normalsize \it 1/AF Bidhannagar, Kolkata 700064, India}\\
\vskip 0.4cm

\vskip 1cm

{\bf ABSTRACT}
\end{center}
Statistical spectroscopic results using the spectral distribution theory 
are obtained for the structure of neutron-rich light nuclei going towards 
the drip line and compared
to experimental values available. These results will be useful for
nuclear astrophysics problems where often averaged nuclear properties 
are adequate.  
\vfill

\noindent $^a$ email: kamales.kar@gmail.com

\section{Introduction}                         \label{sec:introduction}

The study of neutron-rich nuclei, going right upto the drip line, with
experimental results from Radioactive Ion Beam (RIB) has been immensely useful
in the last couple of decades. Explaining many processes in nuclear
astrophysics, like the r-process, needs the nuclear physics inputs of the 
neutron-rich nuclei. On the other
hand the properties of the nuclei far away from stability posed a challenge
to conventional models of theoretical nuclear structure and a deeper 
understanding of the underlying physics. However by now substantial progress
has been made in building new or modifying the existing theoretical models. 
For the lighter nuclei shell model has been successfully used for a detailed 
study of the structure of the very neutron-rich nuclei \cite{Brown-01}. 
It is observed that overall
agreement between theoretical and experimental results are possible with
adequate enlargement of the shell space and the right choice of the two
body interactions and single particle wavefunctions.

Parallel to this, over the last few decades a statistical framework for nuclear
spectroscopy and strength distributions of nuclear excitation and decay 
processes has been developed built on the original prescriptions of the
use of random matrix ensembles in shell model spaces. Many reviews at different
stages of its development and describing its connection to many-body
quantum chaotic systems are available in literature \cite{Brody-81}
\cite{French-82}, \cite{Kota-89}, \cite{Kota-01}, \cite{Kota-10}, \cite{Gomez-11}.
In this theory, often called the spectral distribution theory, statistically 
averaged forms for the nuclear level density
and excitation strength distributions are obtained using arguments of random
matrix ensembles and utilising the group theoretical structure of the
shell model spaces, the averaged quantities are evaluated. These have the
advantage of avoiding explicit diagonalisation of the Hamitonian in many 
particle spaces and need to calculate traces of powers of Hamiltonian as well
as their products with the excitation operators. Though these
methods are suited for nuclei at excitations of a few MeV for a transition
to chaos and the analytical forms obtained in the spectral distribution
methods are asymptotic results, they are seen to
work reasonably well even in the ground state region and for shell model
spaces with not too small number of valence nucleons.

For transition strengths the detailed comparison of the spectral distribution methods with shell model
has been carried out for specific $sd$-shell and $fp$-shell
examples using the same two body interactions for both methods for
electromagnetic and beta decay transitions 
\cite{Draayer-77}, \cite{Kar-81}, \cite{Kota-99}, \cite{Gomez-00} and
\cite{Gomez-11}. For the energy spectra one can go back to a discrete set of energy levels from
the predicted averaged density of states which is continuous in energy and evaluate the
binding energies of nuclei by the Ratcliff prescription \cite{Ratcliff-71}. 
A comparison with the observed binding energies
for the $sd$-shell was done \cite{Chang-71} with the Kuo interaction 
\cite{Halbert-71} and later \cite{Sarkar-87} with the more successful Wildenthal's
Universal $sd$ interaction \cite{Wildenthal-84}. However these were carried out 
demonstrating the success of the spectral distributions
for isotopes around the stable ones and not beyond. Similar applications for
the $fp$ shell nuclei were also done \cite{Kar-97} \cite{Choubey-98} 
 using the interaction KB3 \cite{Poves-81} which has the monopole part
properly adjusted. 
  
These methods are expected to be useful for astrophysical applications as
often the averaged strength functions and level densities are the relevant
quantities there. Spectral distribution methods were earlier used for beta decay and 
electron capture rates during the pre-supernova evolution and subsequent
gravitational collapse leading to supernova explosions \cite{Kar-91} 
\cite{Kar-94} \cite{Kota-95}. 

In this letter we revisit the issue of statistical spectroscopy by spectral
distribution methods in the more challenging region of very
neutron-rich nuclei. We 
describe the calculation of the binding energies of all neutron-rich
nuclei in the $sd$-shell and the evaluation of occupancies and sum rule
strength of a typical electromagnetic transition operator $E2$.
Section 2 will discuss how well the binding energies obtained by 
spectral distribution methods compare with experimental values for nuclei
going close to the drip line. Section 3
is involved in calculating the orbit occupancies and the isoscalar $E2$ sum rule
values demonstrating how spectral distributions can describe the global
features of these quantities over large excitation energies.

\section{Binding Energy}    \label{sec:be}

It is seen that in shell model spaces with large dimensions the smoothed density
of energy eigenvalues goes towards a Gaussian when the number of valence nucleons is not
too small. For the non-interacting case one with a one-body Hamiltonian, one 
gets the result due to the operation of the Central Limit Theorem (CLT) after  
neglecting the Pauli blocking effect. For the interacting case i.e. with (1+2)-body
Hamiltonian, the Gaussian form follows from ensemble averaging using the
Embedded Gaussian Orthogonal Ensemble (EGOE) \cite{Mon-73} \cite{Gomez-11}. 
With $'m'$ valence particles distributed over $'N'$ single particle states and 
with the dimension of the shell model space $d(m)$ (given by $^{N}C_{m}$) the
normalsed density of states $\rho_{m}(E)$ can be expressed in terms of the centroid
$\epsilon(m)$ and width $\sigma(m)$. When the space is partitioned into
subspaces with fixed isospin, to connect to results for real nuclei and with
configuration partitioning, by distributing the $'m'$ particles in $'l'$
shell model orbits giving rise to the normalised configuration-isospin
densities, $\rho_{{\bf m},T}(E)$ go to Gaussians too when each individual 
dimension is large enough. The intensities then just add up.

\be
I_{m, T}(E)= \dis \sum_{\bf m} I_{{\bf m},T} (E)= \dis \sum_{\bf m} d({\bf m},T) \rho_{{\bf m},T} (E)
\ee 

For almost all realistic Hamiltonians the asymptotic Gaussian results holds
and the calculated higher cumulants are seen to be small.  

The ground state energy $E_{g}$ of a nucleus with isospin $T$ and number of 
valence nucleons $m$ is given by the Ratcliff prescription \cite{Ratcliff-71}

\be
\dis \sum_{\bf{m}} \int_{-\infty}^{\bar{E_{g}}} I_{{\bf m},T}(E) dE = d_{0}/2
\ee

where $d_{0}$ is the degeneracy of the ground state. Thus the energy where the 
integrated area below the level density from the low energy side reaches half  
the degeneracy, that value is taken to be the ground state energy.

In this work we consider all nuclei with neutron number equal to or greater than the proton
number, staying within the $sd$-shell i.e. with neutron number not exceeding
20. For locating the ground state energy more accurately we i) consider the
low-lying spectrum of the nuclei and applying the Ratcliff procedure to an
excited state, recover the ground state energy by subtracting the observed excitation
energy from the calculated value, a procedure called `excited state correction' here and 
ii) include corrections due to the small skewness and excess by incorporating 
the Cornish-Fisher expansion \cite{Sarkar-87}. We approximate the third and
fourth moments of Hamiltonian in fixed $(m,T)$ spaces by their scalar i.e.
fixed $(m)$ space values as presently the Spectral Distribution Method (SDM)
codes calculate only the scalar 4th moments. As one goes towards neutron 
drip line nuclei, the experimental excitation spectra are no longer 
available and to conpensate for this an additional isospin dependent
phenomenological correction term needs to be introduced. In  this letter we
take that term as $0.3 T^{2}$. In Table 1 the binding energies 
 for the isotopes of $O$, $F$, $Ne$ and $Na$ calculated using SDM with 
Universal-sd interaction are tabulated and compared to observed
values. As mentioned earlier for the success of these statistical methods 
the dimension of the shell model
spaces should be large and that is why the nuclei having small number 
of valence particles or holes are not included. Table 2 gives the rest of the 
nuclei in the $sd$-shell i.e. isotopes of $Mg$, $Al$, $Si$, $P$, $S$, $Cl$ and 
$Ar$, with their neutron numbers not smaller than the proton numbers. We find  
that overall SDM is able to reproduce the binding energies of the
70 nuclei considered reasonably
well - the average deviation from the experimental values is 0.07 MeV 
i.e. the averaged binding energy value is slightly more than the
observed one and the
RMS deviation is 1.92 MeV. Figure 1 shows the calculated values compared
to the observed ones for the cases of the isotopes $F$ and $Mg$, two typical
nuclei with odd and even proton number. One sees that for nuclei with neutron
number close to or equal to 20 like $^{28}F$, $^{29}Ne$, $^{31}Na$, $^{32}Mg$
and $^{33}Al$ the deviations of the predictions are large
showing the need to enlarge the shell model space for them by including the
lower $fp$-shell orbits like $f_{7/2}$. If one excludes these 5 nuclei the
RMS deviation comes down to 1.65 MeV.


\begin{table}
\begin{tabular}{|c|c|c|c|c|c|}
\hline
Nucleus &Expt. Value& SDM A & SDM B & SDM C  & SDM D     \\
        &   (MeV)   &  (MeV) & (MeV) & (MeV) &   (MeV)   \\
\hline
\hline
$^{21}O$  &  -26.2  &  -29.3  &  -27.5  &  -27.0  &  -24.4   \\
$^{22}O$  &  -32.8  &  -37.8  &  -34.7  &  -33.6  &  -30.0   \\
$^{23}O$  &  -35.3  &  -39.4  &  -39.4  &  -38.0  &  -33.3   \\
$^{20}F$  &  -29.9  &  -33.8  &  -31.1  &  -30.5  &  -29.9   \\
$^{21}F$  &  -37.7  &  -42.2  &  -41.2  &  -39.9  &  -38.8   \\
$^{22}F$  &  -42.6  &  -48.8  &  -46.1  &  -44.9  &  -43.1   \\
$^{23}F$  &  -49.9  &  -55.8  &  -55.8  &  -53.5  &  -50.9   \\
$^{24}F$  &  -53.5  &  -61.9  &  -59.4  &  -57.4  &  -53.8   \\
$^{25}F$  &  -57.6  &  -63.2  &  -62.3  &  -60.8  &  -56.0   \\
$^{26}F$  &  -58.4  &  -66.1  &  -66.1  &  -64.6  &  -58.6   \\
$^{27}F$  &  -59.6  &  -65.0  &  -65.0  &  -64.5  &  -57.0   \\
$^{28}F$  &  -59.2  &  -65.3  &  -65.3  &  -65.2  &  -56.2   \\
$^{20}Ne$ &  -40.9  &  -40.0  &  -38.8  &  -37.4  &  -37.4   \\
$^{21}Ne$ &  -47.1  &  -49.7  &  -47.5  &  -46.0  &  -45.8   \\
$^{22}Ne$ &  -57.3  &  -62.9  &  -59.8  &  -57.4  &  -56.9   \\
$^{23}Ne$ &  -62.1  &  -68.6  &  -66.4  &  -63.8  &  -62.7   \\
$^{24}Ne$ &  -70.7  &  -80.0  &  -76.2  &  -72.6  &  -70.8   \\
$^{25}Ne$ &  -74.7  &  -84.4  &  -81.1  &  -77.8  &  -75.2   \\
$^{26}Ne$ &  -79.8  &  -89.8  &  -87.6  &  -84.0  &  -80.4   \\
$^{27}Ne$ &  -81.0  &  -89.2  &  -89.2  &  -86.7  &  -82.0   \\
$^{28}Ne$ &  -84.6  &  -92.5  &  -92.5  &  -90.5  &  -84.5   \\
$^{29}Ne$ &  -85.6  &  -89.9  &  -89.9  &  -89.5  &  -82.0   \\
$^{22}Na$ &  -58.7  &  -62.3  &  -59.1  &  -57.6  &  -57.6   \\
$^{23}Na$ &  -70.7  &  -76.1  &  -73.3  &  -70.4  &  -70.2   \\
$^{24}Na$ &  -77.3  &  -85.1  &  -82.8  &  -79.5  &  -78.9   \\
$^{25}Na$ &  -86.0  &  -97.5  &  -94.1  &  -89.6  &  -88.4   \\
$^{26}Na$ &  -91.2  & -101.7  &  -99.0  &  -95.3  &  -93.5   \\
$^{27}Na$ &  -97.6  & -107.3  & -107.3  & -103.0  & -100.4   \\
$^{28}Na$ & -100.8  & -112.2  & -112.2  & -108.4  & -104.8   \\
$^{29}Na$ & -110.2  & -113.5  & -113.5  & -111.0  & -106.2   \\
$^{30}Na$ & -107.0  & -113.7  & -112.1  & -111.1  & -105.1   \\
$^{31}Na$ & -110.5  & -113.9  & -113.9  & -113.5  & -106.0   \\

\hline
\end{tabular}
\caption{
Calculated binding energies by Spectral Distribution Methods (SDM) 
of isotopes of $O$, $F$, $Ne$ and $Na$ 
with neutron number equal or more than the proton number, compared
to experimental values. The experimental values are with respect to $^{16}O$ 
ground state and with the Coulomb part subtracted. `A' refers to the
Ratcliff procedure for only the ground state, `B' refers to the ground state
with excited state correction, `C' refers to the one with excited state as
well as non-zero skewness and excess and finally `D' is the one which includes
the isospin dependent correction to the values in column `C'.
}
\end{table}

\begin{table}
\begin{tabular}{|c|c|c|c|c|c|}
\hline
Nucleus &Expt. Value& SDM A & SDM B & SDM C  & SDM D    \\
        &   (MeV)   &  (MeV) & (MeV) & (MeV) &   (MeV)  \\
\hline
\hline
$^{24}Mg$  &  -87.5  &  -93.1  &  -88.9  &  -85.0  &  -85.0   \\
$^{25}Mg$  &  -94.4  & -101.8  &  -98.7  &  -94.5  &  -94.3   \\
$^{26}Mg$  & -105.1  & -116.6  & -110.7  & -105.3  & -104.7   \\
$^{27}Mg$  & -111.2  & -123.4  & -119.7  & -114.2  & -113.1   \\
$^{28}Mg$  & -119.2  & -131.3  & -127.1  & -121.8  & -120.0   \\
$^{29}Mg$  & -122.6  & -132.6  & -131.1  & -127.3  & -124.7   \\
$^{30}Mg$  & -128.6  & -138.4  & -134.8  & -132.2  & -128.6   \\
$^{31}Mg$  & -130.6  & -138.5  & -135.9  & -134.8  & -130.1   \\
$^{32}Mg$  & -136.2  & -140.1  & -137.6  & -137.3  & -131.3   \\
$^{26}Al$  & -106.1  & -113.7  & -110.4  & -106.6  & -106.6   \\
$^{27}Al$  & -118.7  & -127.8  & -124.0  & -119.0  & -118.8   \\
$^{28}Al$  & -126.0  & -137.2  & -133.6  & -127.9  & -127.3   \\
$^{29}Al$  & -135.0  & -145.3  & -143.1  & -138.2  &  137.1   \\
$^{30}Al$  & -140.3  & -150.8  & -148.1  & -144.7  & -142.9   \\
$^{31}Al$  & -147.0  & -156.2  & -156.2  & -149.1  & -146.5   \\
$^{32}Al$  & -150.9  & -159.6  & -155.7  & -154.7  & -151.1   \\ 
$^{33}Al$  & -156.1  & -159.5  & -157.2  & -156.1  & -152.2   \\
$^{28}Si$  & -136.0  & -145.4  & -141.4  & -134.9  & -134.9   \\
$^{29}Si$  & -143.9  & -155.3  & -150.4  & -144.8  & -144.6   \\
$^{30}Si$  & -154.1  & -167.3  & -162.8  & -157.0  & -156.4   \\
$^{31}Si$  & -160.2  & -169.7  & -166.7  & -162.9  & -161.8   \\
$^{32}Si$  & -169.0  & -178.2  & -175.6  & -172.0  & -170.2   \\
$^{33}Si$  & -173.0  & -179.0  & -177.8  & -176.2  & -173.6   \\
$^{34}Si$  & -180.2  & -184.5  & -183.1  & -182.1  & -178.6   \\
$^{30}P$   & -155.5  & -166.5  & -161.0  & -156.9  & -156.9   \\
$^{31}P$   & -167.3  & -178.5  & -173.3  & -169.1  & -168.8   \\
$^{32}P$   & -174.7  & -185.9  & -181.3  & -178.2  & -177.6   \\
$^{33}P$   & -184.4  & -193.4  & -189.5  & -187.1  & -186.0   \\
$^{34}P$   & -190.2  & -197.7  & -194.9  & -193.7  & -191.9   \\
$^{35}P$   & -198.1  & -202.2  & -200.6  & -200.1  & -197.4   \\
$^{32}S$   & -182.4  & -191.6  & -186.1  & -182.5  & -182.5   \\
$^{33}S$   & -190.5  & -198.4  & -195.8  & -193.2  & -192.9   \\
$^{34}S$   & -201.4  & -209.6  & -205.5  & -203.3  & -202.7   \\
$^{35}S$   & -207.9  & -213.2  & -211.3  & -210.3  & -209.2   \\
$^{34}Cl$  & -202.2  & -212.4  & -204.9  & -203.8  & -203.8   \\
$^{35}Cl$  & -214.3  & -219.4  & -216.3  & -215.4  & -215.2   \\
$^{36}Cl$  & -222.3  & -227.2  & -224.4  & -224.2  & -223.6   \\
$^{36}Ar$  & -229.6  & -232.9  & -230.9  & -230.1  & -230.1   \\

\hline
\end{tabular}
\caption{
Calculated binding energies by SDM 
of isotpoes of $Mg$, $Al$, $Si$, $P$, $S$, 
$Cl$ and $Ar$ with neutron number equal to or more than the 
proton number, compared
to experimental values. For details of the experimental values and  
the SDM columns `A' to `D'
see caption of Table 1. 
}

\end{table}

\begin{figure}
\includegraphics[width=.70\columnwidth,angle=270]{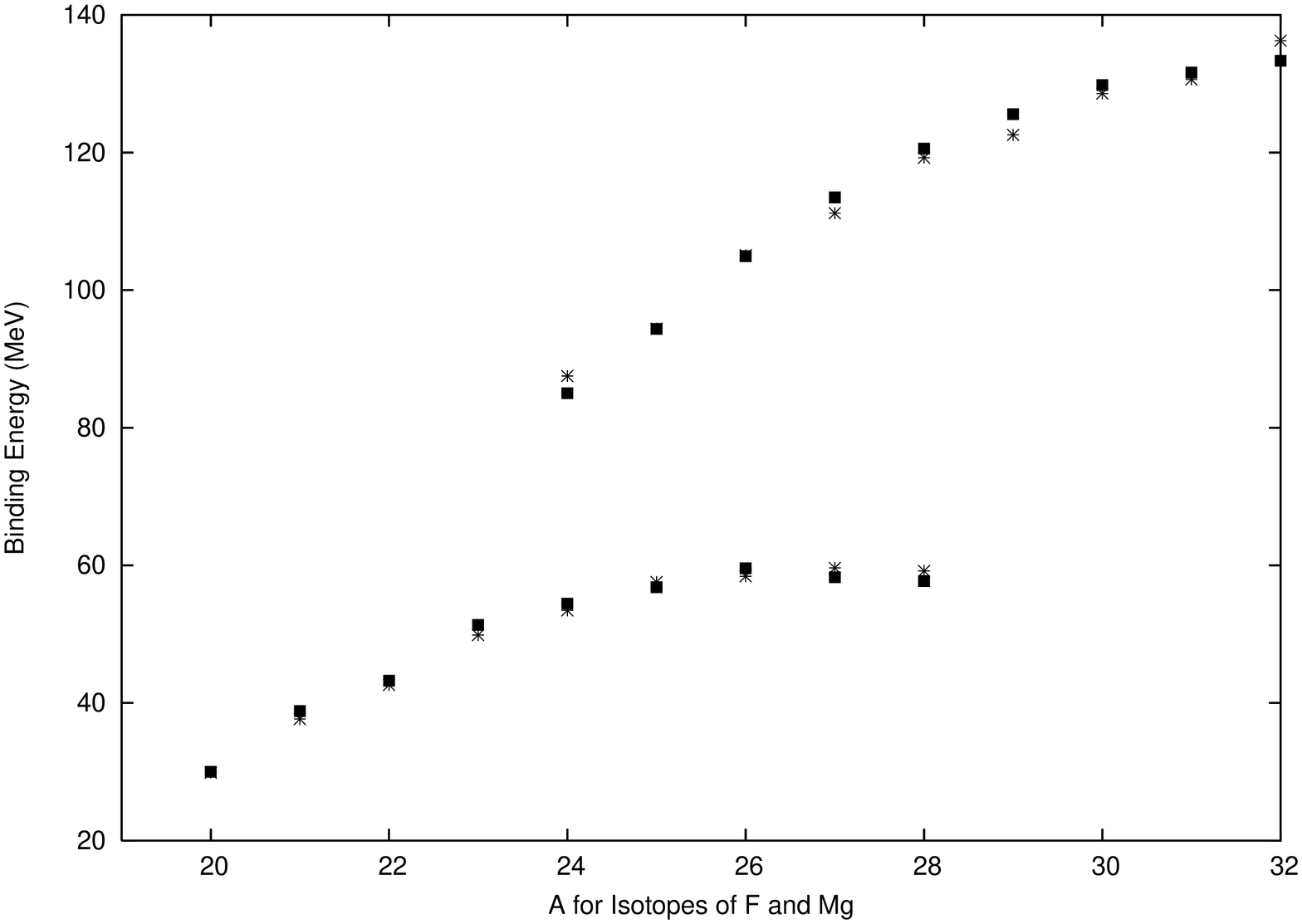}
\caption{\label{fig:be3}
The absolute values of the binding energies of isotopes of F and Mg by spectral 
distribution methods compared to experimental values. The stars stand for the
experimental values and the filled squares for the spectral distribution 
predictions. All the values are with respect to $^{16}O$ ground state energy and
with the Coulomb contribution subtracted.
}
\end{figure}

We also compare the predictions of binding energies using the A-dependent two 
body interaction Universal-sd with another two body interaction, 
Chung-Wildenthal (CW) \cite{Vary-77}.
Figure 2 gives the calculated corrected binding energies for
all the neutron-rich isotopes considered for the element $Ne$ for both 
interactions alonwith the experimental values. One sees that as one goes away
from stability Universal-sd does a better job than CW, as expected.


\begin{figure}
\includegraphics[width=.70\columnwidth,angle=270]{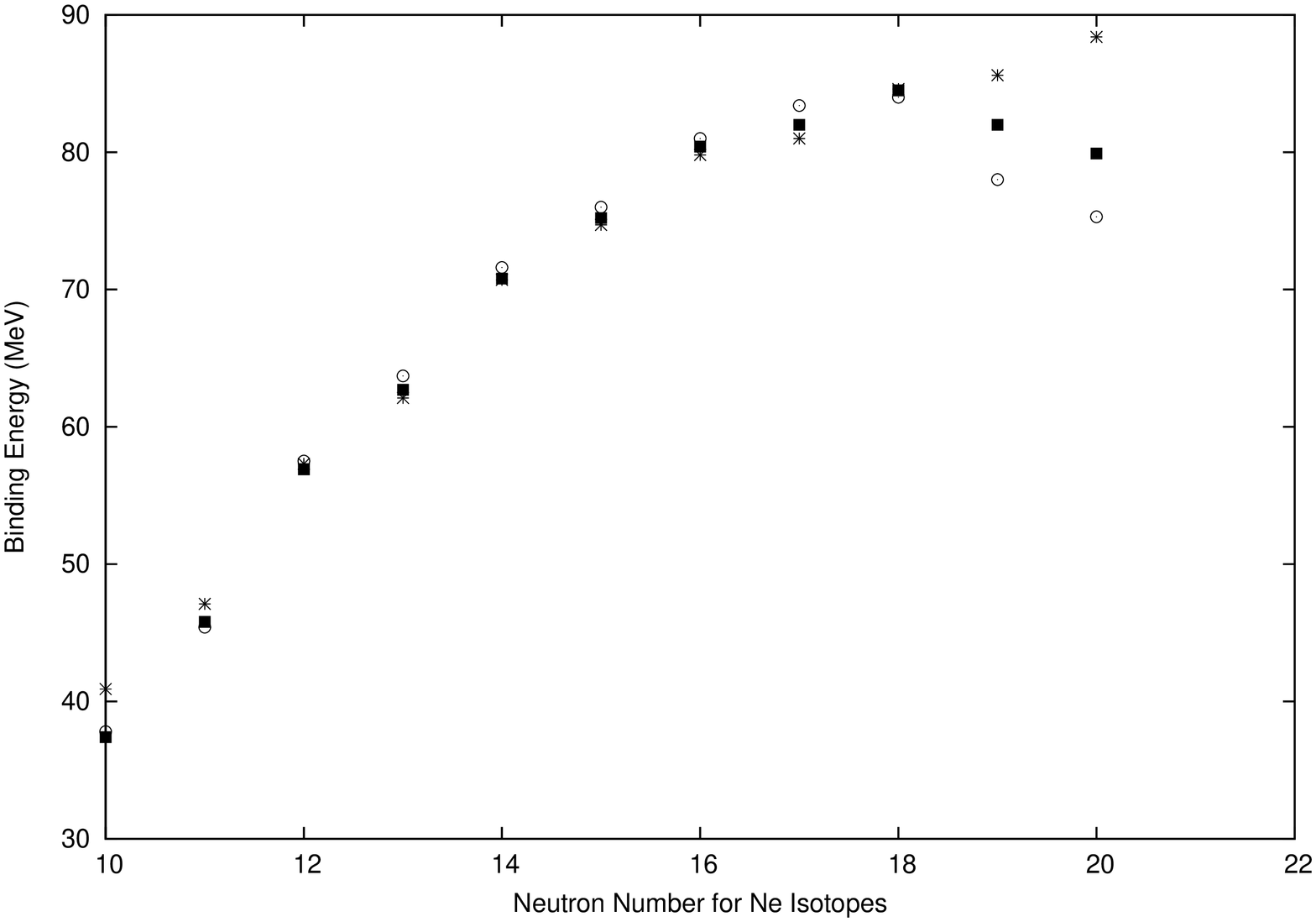}
\caption{\label{fig:be4}
The absolute values of the binding energies of isotopes of $Ne$ by the 
two interactions Universal-sd
and CW compared to experimental values. Stars stand for the experimental
numbers whereas the filled squares and the empty circles stand for the 
values for Universal-sd and CW interactions respectively.
}
\end{figure}

\section{Excitation Strength Sum Rule and Orbit Occupancy}

Taking the density of energy eigenvalues $\rho (E)$ as the weight function one
can define a unique set of orthogonal polynomials $P_{\mu}(E)$. If the density 
is Gaussian, the polynomials are Hermite. The expectation value of an operator
$K$ in the energy eigenstate $ |E>$ is given by

\be
 \lan E|K|E \ran =\dis \sum_{\mu} \lan KP_{\mu}(H) \ran^{m} P_{\mu}(E)= 
\lan K \ran^{m} +  \zeta_{K,H} (\sigma_{K}) (E-\epsilon_{1})/\sigma_{1}+\dots
\ee

Keeping the first two terms above is called the CLT result and is seen to to be
 true when the spectrum of the eigenvalues of H is a Gaussian and that remains
so under the transformation $H \rightarrow H+ \alpha K $ for small $\alpha$ 
\cite{Chang-73}. This gives the expectation value of the operator $K$ 
as a function of the energy E a geometric
interpretation in terms of the correlation coefficient $\zeta_{K,H}$. 
if $K=O^{\dagger} O$ with $O$ a one-body excitation or decay operator,
then its expectation value in the state $|E>$ gives the sum rule rule
strength of excitation by the operator $O$. With the correlation 
coefficient $\zeta_{K,H}$ 
negative and large, the ground state region has much larger averaged strength
sum than the high excitation region. On the other hand, if $\zeta_{K,H}$
has a large positive value, then the ground state region has much smaller
sum rule strength than the region with high excitation energy.

The configuration-averaged expression for the expectation value of
operator $K$ in the energy eigenstate $|E>$ for the CLT case, is 

\be
\tilde K(E)_{CLT}= \dis \sum_{\bf m} [I_{{\bf m}, T}(E)/I_{m,T}(E)] 
[\lan K \ran^{{\bf m},T}+\tilde \zeta_{K,H} \tilde \sigma_{K}(E-\tilde \epsilon_{1})/\tilde \sigma_{1}]
\ee 

where $\tilde \zeta_{K,H}$ is the correlation coefficient between operators
$K$ and $H$ in the configuration-isospin space $({\bf m},T)$ alongwith
$\tilde \sigma_{K}$, $\tilde \epsilon_{1}$ and $\tilde \sigma_{1}$ 
the width of $K$, centroid of $H$ and width of $H$ in the configuration-isospin
space respectively.

\begin{figure}
\includegraphics[width=.70\columnwidth,angle=270]{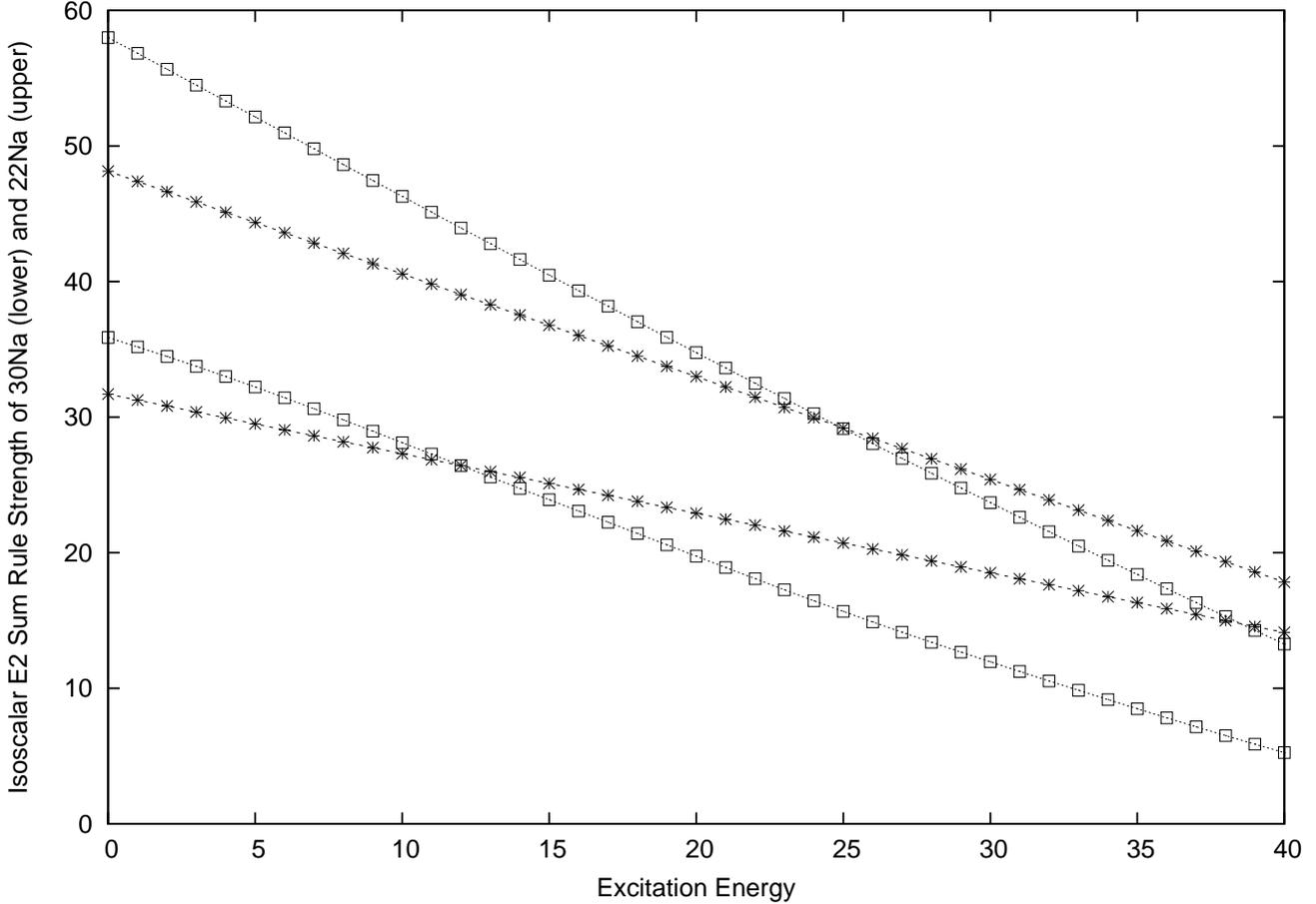}
\caption{\label{fig:qqna}
The sum rule strength for isoscalar E2 excitation as function of excitation
energy for the nuclei $^{30}Na$ (the lower lines)  and $^{22}Na$ (the
upper lines) calculated by spectral distribution theory. For both nuclei
the line with stars stands for the scalar result and 
the line with unfilled squares for the configuration result. The isoscalar
E2 sum rule operator is normalized by the factor $(2.17)^2$ as done in
\cite{Draayer-77}.
}
\end{figure} 

To illustrate this we take the example of isoscalar E2 excitation for
sd shell nuclei. Earlier one saw that the correlation coefficients of isoscalar
E2 sum rule operator with Kuo interaction \cite{Kuo-67} with $^{17}O$ single
particle energies \cite{Kuo-67} for a typical
example of 6 particles in sd-shell are -0.52, -0.50, -0.46 and -0.36 
for spaces with 
T=0,1,2 and 3 respectively \cite{Draayer-77}. We find that for the Universal-sd
interaction the correlation coefficients of isoscalar E2 are -0.48 and 
-0.36 for the isotopes $^{22}Na$ (with T=0) and $^{30}Na$ (with T=4)
respectively. Fig. 3 shows that the CLT result for the averaged isosclar E2 sum
rule strength decreases with excitation energy for the nucleus $^{22}Na$
with T=0 states as well as for the neutron-rich nucleus $^{30}Na$ with
T=4 states. However the decrease is faster for $^{22}Na$ than for 
$^{30}Na$. Thus this type 
of analysis for different excitation operators help us to understand the
global features of the strength sum and strength distributions. 

If one takes $K$ as $n_{r}$, the occupancy in the shell model orbit `r', one 
gets an even simpler form for the oocupancy 

\be
n_{r}(E)=\dis\sum_{\bf m} [I_{{\bf m},T}(E)/I_{m,T}(E)] [m_{s}({\bf m},T)]
\ee

where $m_{r}({\bf m},T)$ stands for the occupancy in the confiuration-isospin
space $({\bf m},T)$.

Thus using SDM one can easily evaluate the occupancy of all the three orbits in
sd-shell for the ground state as well as the excited states.
 Fig. 4 gives the three ground state occupancies for the example of the
isotopes of the elements $F$ and $Mg$ as functions of increasing mass number,  
$A$. One observes in the figure that though with the increasing value of the  
number of valence nucleons all three occupancies increase for both $F$ and 
$Mg$, but the orbit $d_{5/2}$ show faster increase than the other two in the
lower half of the shell. When the neutron number nears the value of 20,
$d_{5/2}$ occupancy shows a decrease, stressing the need to include
fp-shell orbits (at least the lowest $f_{7/2}$ orbit) in the calculation. 

In conclusion, we demonstrate in this letter that the spectral distribution
theory works reasonably well to reproduce global structural properties of very
neutron-rich nuclei in the sd shell.   
This avoids the matrix diagonalisation involved in shell model and is much 
simpler in computation and in the interpretation of the global features than 
other models of nuclear structure. We plan to extend this to other excitation 
operators for understanding the strength distributions and sum rule strengths
as well as apply this to heavier nuclei.

\begin{figure}
\centerline{\includegraphics[width=.35\columnwidth,angle=270]{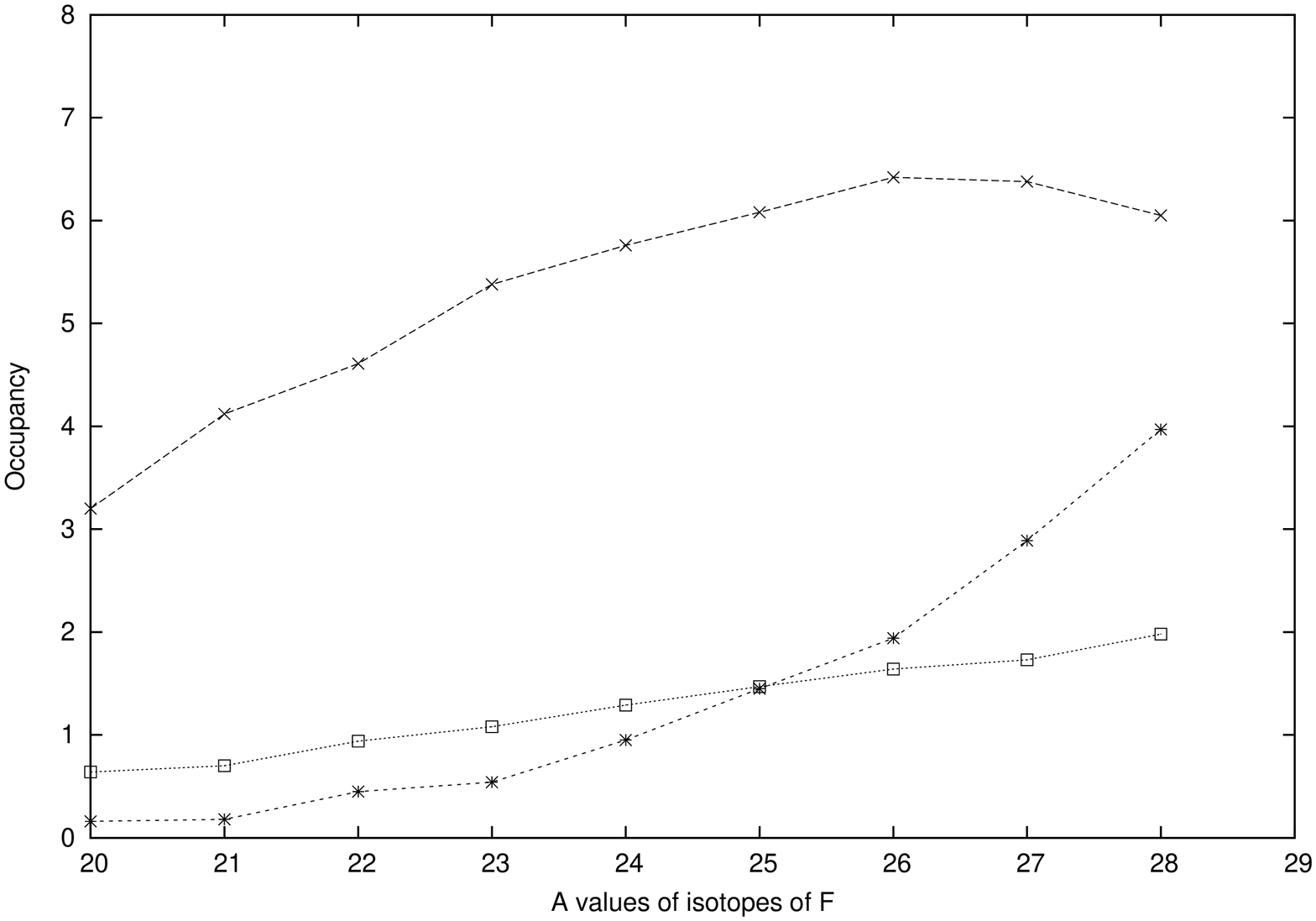}\includegraphics[width=.35\columnwidth,angle=270]{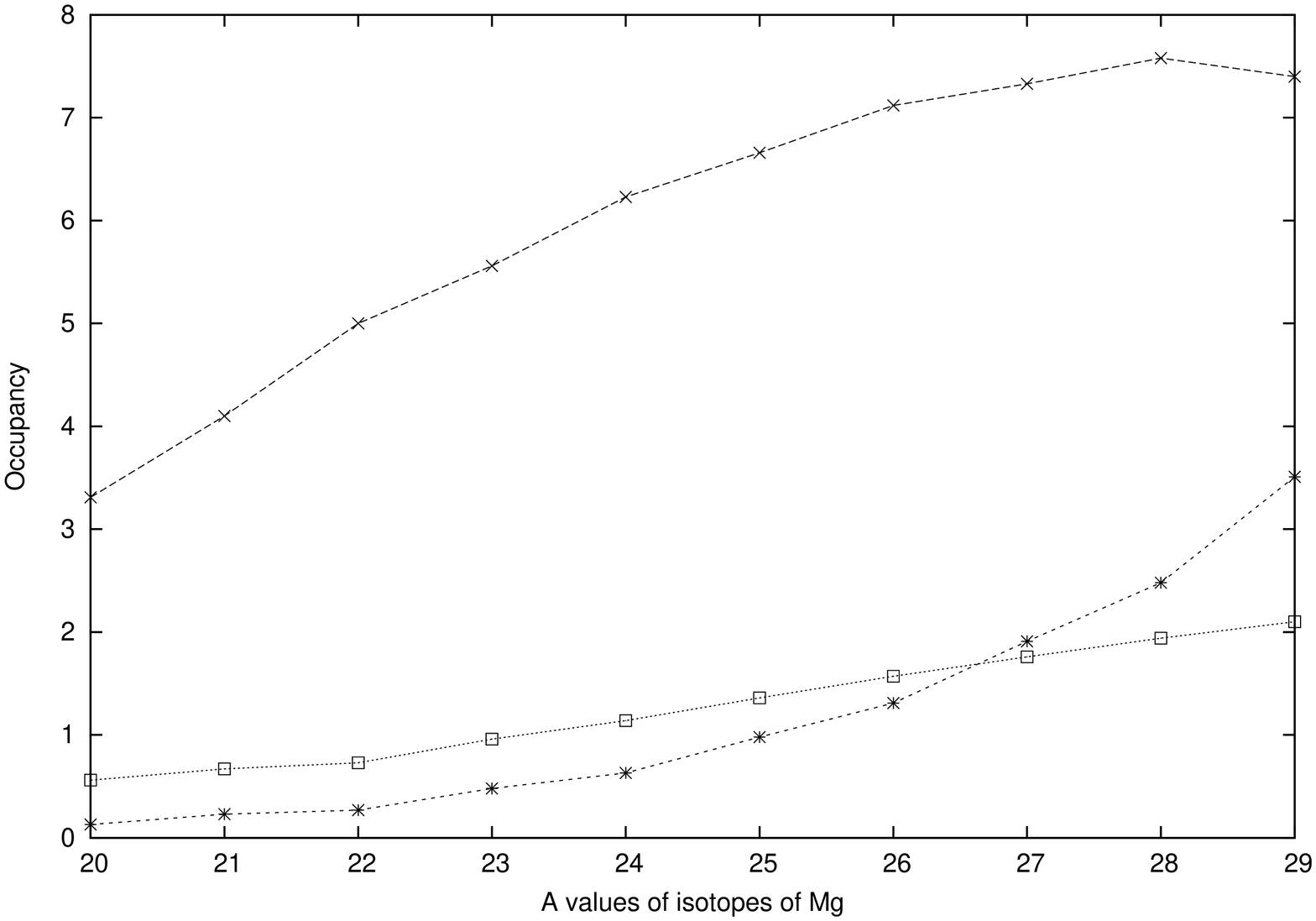}}
\caption{\label{fig:occp1}
The predicted occupancy of isotopes of $F$ and $Mg$ for the three shell model
orbits $d_{5/2}$, $d_{3/2}$ and $s_{1/2}$.
}
\end{figure}

The author acknowledges helpful discussions with V.K.B. Kota.


\begin{thebibliography}{99}

\bibitem{Brown-01}
B.A. Brown, Prog. Part. Nucl. Phys. {\bf 47} 517 (2001)

\bibitem{Brody-81}
T.A. Brody, J. Flores, J.B. French, P.A. Mello, A. Pandey and S.S.M. Wong,
Rev. Mod. Phys. {\bf 53} 385 (1981)

\bibitem{French-82}
J.B. French and V.K.B. Kota, Ann. Rev. Nucl. Part. Sci. {\bf 32} 35 (1982)

\bibitem{Kota-89}
V.K.B. Kota and K. Kar, Pramana- J. Phys. {\bf 32} 647 (1989)

\bibitem{Kota-01}
V.K.B. Kota, Phys. Rep. {\bf 347} 223 (2001)

\bibitem{Kota-10}
V.K.B. Kota and R.U. Haq, Spectral Distributions in Nuclei and Statistical
Spectroscopy, World Scientific, Singapore, 2010

\bibitem{Gomez-11}
J.M.G. Gomez, K. Kar, V.K.B. Kota, R.A. Molina, A. Relano and J. Retamosa,
Phys. Rep. {\bf 499} 103 (2011)

\bibitem{Draayer-77}
J.P. Draayer, J.B. French and S.S.M. Wong, Ann. Phys. (N. Y.) {\bf 106} 503 (1977)

\bibitem{Kar-81}
K. Kar, Nucl. Phys. A {\bf 368} 285 (1983)

\bibitem{Kota-99}
V.K.B. Kota, R. Sahu, K. Kar, J.M.G. Gomez and J. Retamosa Phys. Rev. C {\bf 60}
 051306 (1999)

\bibitem{Gomez-00}
J.M.G. Gomez, K. Kar, V.M.Manfredi, R.A. Molina and J. Retamosa, Phys. Lett. B
{\bf 480} 245 (2000)

\bibitem{Ratcliff-71} K.F. Ratcliff, Phys. Rev. C {\bf 3} 117 (1971)

\bibitem{Chang-71}
F.S. Chang, J.B. French and T.H. Thio, Ann. Phys. ( N.Y.) {\bf 66} 137 (1971)

\bibitem{Halbert-71}
E.C. Halbert, J.B. McGrory, B.H. Wildenthal and S.P. Pandya, Adv. Nucl. Phys.
{\bf 4} 315 (1971)

\bibitem{Sarkar-87} S. Sarkar, K. Kar and V.K.B. Kota, Phys. Rev. C {\bf 36} 
2700 (1987)

\bibitem{Wildenthal-84}
B.H. Wildenthal, Prog. Part. Nucl. Phys. {\bf 11} 5 (1984)

\bibitem{Kar-97}
K. Kar, S. Sarkar, J.M.G. Gomez, V.M. Manfredi and L. Salasnich, Phys. Rev. C
{\bf 55} 1260 (1997)

\bibitem{Choubey-98} S. Choubey, K. Kar, J.M.G. Gomez and V.R. Manfredi,
Phys. Rev. C {\bf 58} 597 (1998)

\bibitem{Poves-81} A. Poves and A.P. Zuker, Phys. Rep. {\bf 70} 235 (1981)

\bibitem{Kar-91}
K. Kar, S. Sarkar and A. Ray, Phys. Lett. B {\bf 261} 217 (1991)

\bibitem{Kar-94} K. Kar, A. Ray and S. Sarkar, Ap. J. {\bf 434} 662 (1994)

\bibitem{Kota-95} V.K.B. Kota and D. Majumdar, Z. Phys. A {\bf 351} 377 (1995)

\bibitem{Mon-73}
K.K. Mon and J.B. French, Ann. Phys. (N. Y.) {\bf 78} 111 (1973)

\bibitem{Vary-77} J.P. Vary and S.N. Yang, Phys. Rev. C {\bf 15} 1545 (1977)

\bibitem{Chang-73} F.S. Chang and J.B. French, Phys. Lett. B {\bf 44} 131 (1973)

\bibitem{Kuo-67} T.T.S. Kuo, Nucl. Phys. A {\bf 103} 71 (1967)

\end{thebibliography}
\end{document}